\documentclass[12pt]{iopart}

\usepackage{bm}
\usepackage{amssymb}
\usepackage{amsfonts}
\usepackage{graphicx}

\begin{document}

\title{Direct WIMP identification: Physics performance of a segmented noble-liquid target immersed in a Gd-doped water veto}

\author{A Bueno, M C Carmona and A J Melgarejo}

\address{Dpto. de F{\'\i}sica Te\'orica y del Cosmos \& C.A.F.P.E., 
Universidad de Granada, Spain}
\eads{\mailto{a.bueno@ugr.es}, \mailto{carmencb@ugr.es}, \mailto{amelgare@ugr.es}}

\begin{abstract}
We evaluate background rejection capabilities and physics performance of 
a detector composed of two diverse elements: a sensitive target (filled 
with one or two species of liquefied noble gasses) and an active veto (made of 
Gd-doped ultra-pure water). A GEANT4 simulation shows that for 
a direct WIMP search, this device can reduce the neutron background to 
O(1) event per year per tonne of material. Our calculation shows that an exposure 
of one tonne $\times$ year will suffice to exclude spin-independent 
WIMP-nucleon cross sections ranging from $10^{-9}$ pb to $10^{-10}$ pb.
\end{abstract}

\maketitle

\section{Introduction}

Evidence for the existence of dark matter is overwhelming.
The first was found by F. Zwicky in 1933
while measuring the rotational velocity of the Coma cluster of
galaxies~\cite{Zwicky}. Since then astronomers have gathered 
more relevant data, which likely point in the direction of the existence
of a new form of matter~\cite{Trimble}. For instance, the striking optical and X-ray
images, obtained by the Chandra telescope, of the so-known bullet cluster
(1E0657-558)~\cite{bullet,bullet2} cannot be explained simply by advocating that Newtonian 
dynamics is modified at large scales~\cite{Sanders}. Additional support
for dark matter comes from the precise measurement of the cosmic
microwave background radiation done by the Wilkinson Microwave Anisotropy Probe 
(WMAP). Together with
the Sloan Digital Sky Survey (SDSS) large-scale structure data, it tells us that only a small
fraction of the matter content of the Universe ($\Omega_B = 0.042\pm 0.002$) is of
baryonic origin, while the rest is made of a totally unknown new form of
matter ($\Omega_{DM} = 0.20\pm 0.02$)~\cite{wmap}. The rest of the
energy content of the Universe ($\Omega_\Lambda = 0.76\pm 0.02$) is
accounted for by a very smooth form of energy called dark energy.

So far dark matter has been observed only through its gravitational
effects, therefore we know little about its fundamental
properties. Massive neutrinos contribute to a small fraction of dark
matter~\cite{atm,solar}. However the most plausible hypothesis for potential candidates 
(generically known as Weakly Interacting Massive Particles, WIMPs) states that they 
should be neutral weakly-interacting heavy particles with lifetimes 
comparable to the age of the Universe. No such thing exists in the
Standard Model of Particle Physics. However theories beyond 
it (like, for example, Supersymmetry, Extra Dimensions, etc.) do have 
particles that naturally arise as dark matter candidates~\cite{Bertone}.

Parallel to model building developments, there is an intense
and challenging experimental activity devoted to WIMP detection, 
as well~\cite{Gaitskell}. DAMA/NaI claims 
to have found evidence for the presence of WIMPs in the
Galactic halo~\cite{Bernabei:2000qi}. Using a direct detection method, they have measured an 
annual modulation, over seven annual cycles (107,731 kg day total
exposure), consistent with expectations from a WIMP signature. 
The collected DAMA/LIBRA data~\cite{Bernabei:2008yi} supports this claim. 
Considering the data from both experiments (amounting to an 
exposure of 0.82 tonnes $\times$ yr), the presence of dark matter particles 
in the galactic halo is supported at 8.2 $\sigma$ C.L. However, 
the situation is highly controversial since other
direct-search experiments, probing similar regions of the parameter space, have
found negative
results~\cite{Akerib:2004fq,Chardin:2003vn,Ahmed:2003su,warp,xenon10}. 
Indirect detection methods have not found signals that could be
attributed to WIMP annihilation~\cite{Ambrosio:1998qj,Desai:2004pq}. 

To improve current sensitivities and 
explore in depth the parameter space of the most favoured dark matter models,
there is an indisputable need for more massive detectors with enhanced background
rejection capabilities. The use of
liquefied noble gasses, as target for WIMP interactions, ranks
among the most promising detection techniques~\cite{nobleliquids1,nobleliquids2,nobleliquids3}. This technology is easily scalable and
allows to build detectors in the range of few tonnes of fiducial
mass~\cite{andre}. This paper concentrates on evaluating the physics
potential of one of such detectors. An effort has been made to design
an experiment that allows to reduce as much as possible the background caused by neutron
interactions inside the active target. In addition, we consider several independent target units
that can be filled with different noble liquids; in this way we can 
aim at confirming, with a single experiment, that 
the event rate and the recoil spectral shape follow 
the expected dependence on A$^2$ for WIMP signals~\cite{pfsmith}. 
In the following Sections, we
describe the basic detector layout and its foreseen physics performance. 

\section{Detector description}
\label{sec:det}

Liquid noble elements, used as sensitive medium for direct dark
matter searches, are a promising alternative to ionization, solid
scintillation and milli-Kelvin cryogenic detectors. When a WIMP
particle scatters off a noble element, scintillation photons
and ionization electrons are produced due to the interactions of the
recoiling nucleus with the neighbouring atoms. The simultaneous
detection of primary scintillation photons and ionization charge (or
the secondary photons produced when this charge is extracted from the
liquid to the gas phase~\cite{rusos}) is a powerful discriminator against
backgrounds. Pulse shape provides an additional tool 
to identify true signals: depending on 
the nature of the interacting particle, the scintillation light 
shows a different time dependence~\cite{Hitachi}.  
In addition, thanks to the high level of purity achieved, 
these detectors can drift ionization charges for several meters, 
hence it is conceivable to reach masses of the order of several tonnes. 
Nowadays XENON~\cite{xenon10}, ZEPLIN~\cite{zeplin1,zeplin2,zeplin3,zeplin4}, 
XMASS-DM~\cite{xmass}, WARP~\cite{warp} and ArDM~\cite{andre} collaborations 
use liquid argon or xenon targets to look for WIMPs. Similar detectors
can be used to detect the yet unobserved coherent
neutrino-nucleus elastic scattering~\cite{betapaper}.

Assuming a one tonne detector, we expect an event rate of
O(10) events per year of operation for a WIMP-nucleus cross section 
of 10$^{-10}$ pb. To explore such small cross sections, 
backgrounds should be reduced to very challenging levels (about 
1 event per tonne per year). In case the target is filled with argon, 
the $^{39}$Ar isotope, which is a beta particle emitter, is a serious source 
of concern (its activity is approximately 1~Bq per kg of natural
argon~\cite{Ar39radio}). However, the most important source of
background is due to neutrons produced in detector components 
or in the rock of the underground cavern. A large fraction of these
external neutrons can be rejected using external hydrocarbon
shields, active vetoes or a combination of the two~\cite{cline}. 
High-energy neutrons induced by muon
interactions in the rock are a more serious concern. Recently, an innovative
neutron multiplicity meter, very similar in concept to the detector
discussed in this document, has been proposed to monitor this neutron
flux~\cite{akerib}. The flux of internal neutrons can be highly 
reduced using low activity materials for the inner parts of the
detector. However, it is unavoidable that some of them interact with 
target nuclei mimicking a WIMP signal. 

Detectors with a large fiducial volume offer the
advantage of an increased probability for neutrons to interact
several times, before they exit the target. For WIMPs this is highly unlikely
given the small cross sections involved. This fact can be used to 
further reduce neutron backgrounds. In our case, given the reduced dimensions 
of the sensitive targets (see below), the signal due to multiple interactions cannot 
be the main tool for background rejection. To reduce neutron contamination, 
we propose a detector made of two diverse elements: 
an external detector, acting as an active veto, made of ultra-pure water
doped with gadolinium. This will enhance neutron capture and its
posterior identification~\cite{gadzooks}. Immersed in this veto, 
the internal detector, that acts as target, consists of cells filled
with a noble element. This configurations allows to operate the 
sensitive target with two noble liquids simultaneously. 
If energy deposits occur,
within a certain time window, both in the cell and water,
the event is tagged as neutron-like provided the external veto records the
typical 8 MeV gamma cascade from neutron capture on gadolinium.

\subsection{Noble liquid target}
\label{sec:target}

\begin{figure}[htb]
 \centering
 \includegraphics[scale=0.3]{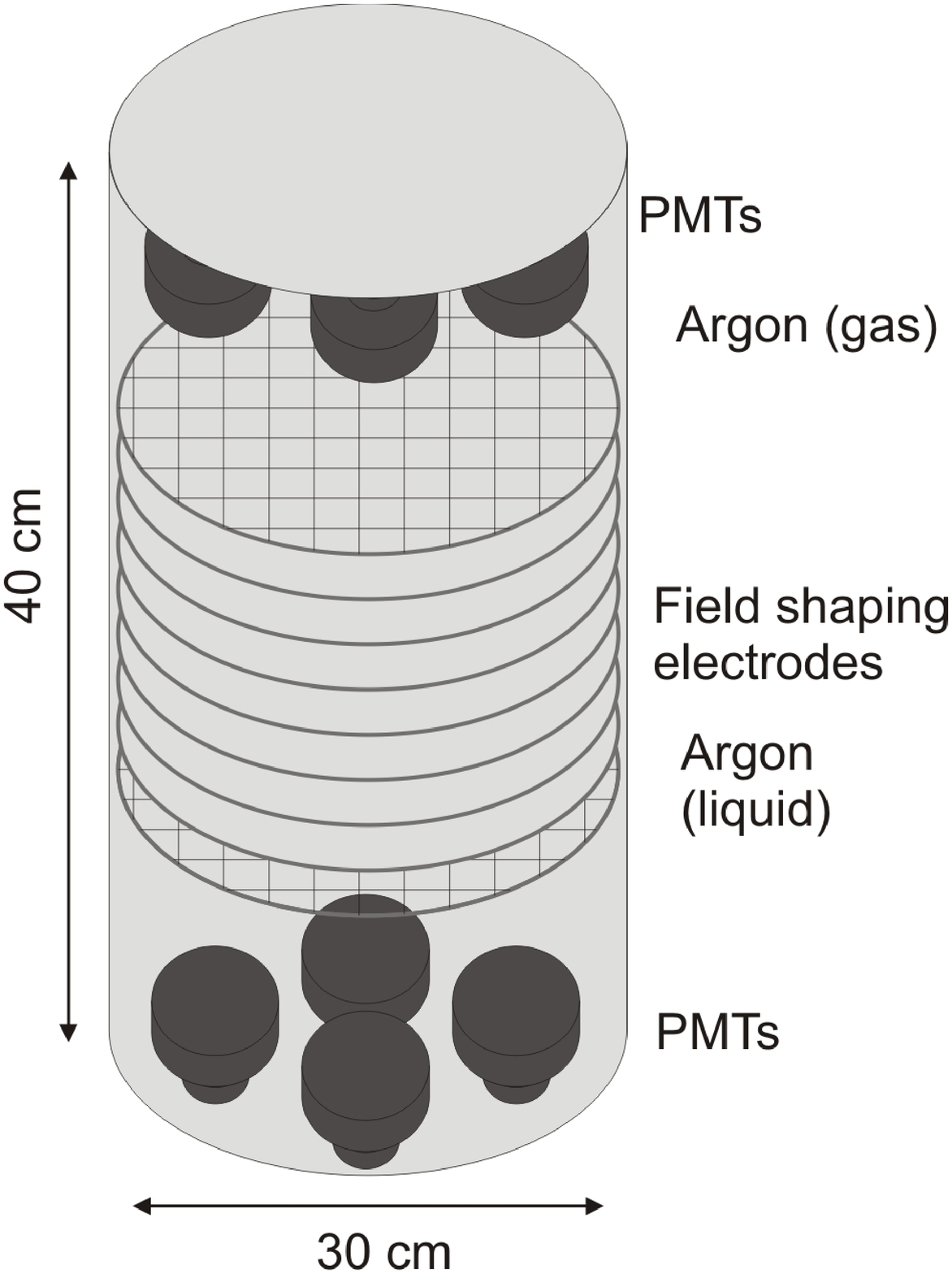}\\
 \includegraphics[scale=0.5]{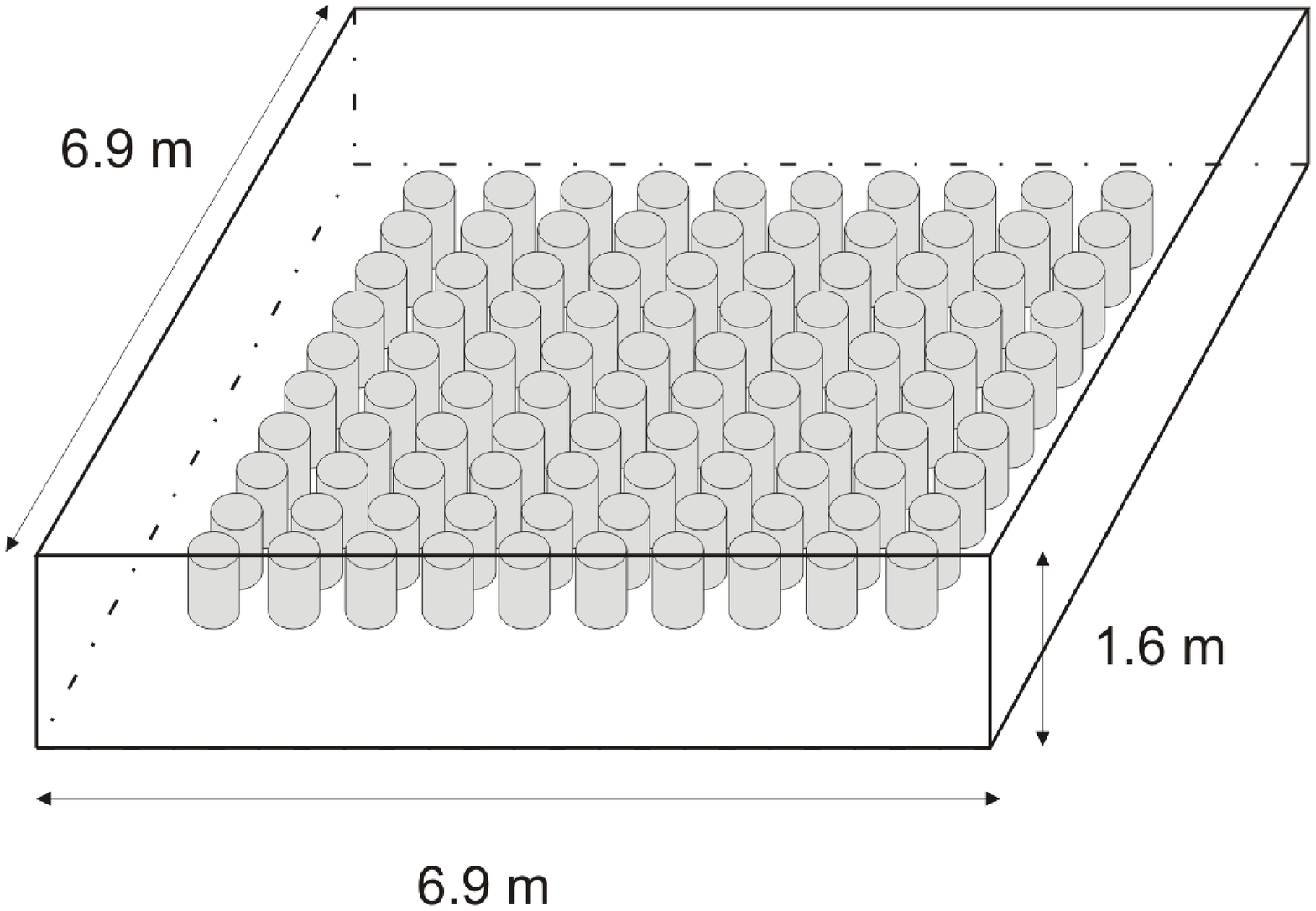}\\
 \caption{
  Artist's view of the detector: (Top) A target cell. (Bottom) Noble liquid target 
plus active veto.}
 \label{fig:detsim}
\end{figure}

We have carried out a full simulation of the detector using 
GEANT4~\cite{Allison} (see Figure~\ref{fig:detsim}). The 
target is made of 100 low-background metal cylinders 
(each 40 cm high and 30 cm in diameter). The internal 
volume, that can be filled with a noble liquid, has 30 cm drift 
distance and 24 cm in diameter. For our physics 
studies, the fiducial region corresponds to a cylinder of 
6 cm radius and 25 cm high. The fiducial mass amounts up to 0.8
tonnes in case the target is filled with liquid xenon (LXe) and 0.4
tonnes in case liquid argon (LAr) is used. Our device can
detect simultaneously the ionization charge and the scintillation
light resulting from the scattering of incoming particles off xenon or
argon nuclei. Light is read by means of photomultiplier tubes (PMTs) 
placed at the target bottom. 
Ionization electrons are drifted to the liquid surface where
they are  converted into secondary scintillation light that is read by
PMTs on top of the cylinders. Charge amplification devices (i.e., GEM, LEM,
Micromegas~\cite{Sauli,Jeanneret,Giomataris2}) are a possible
alternative for charge readout.  

This configuration of independent cylinders, apart from the fact of
being easily scalable, offers a clear experimental advantage: 
data taking can proceed with two different targets
simultaneously. Cylinders can be filled with argon and xenon, for
example. In case a WIMP signal is observed with enough statistical
relevance, we can confirm in a single experiment that the event rate 
and the recoil spectral shape follow the expected
dependence on A$^2$. 

\subsection{Water-\v{C}erenkov neutron detector}
\label{sec:veto}

The active target is immersed in a water tank (1.6~m height, 6.9~m width
and 6.9~m long), made of copper or other
low background material. The distance between cylinders is 30~cm. 
The distance to the veto walls is 60~cm. This
distance has been optimized to allow for an efficient neutron capture by
gadolinium. The veto contains 70 tonnes of ultra-pure water, once 
we subtract the volume taken by the sensitive targets and the ancillary system.  
1500 9" PMTs (40\% photo-coverage), mounted at the water-tank
walls, are used to detect 
the photons produced by neutron capture on Gd. They will detect 
the light produced by penetrating cosmic muons, as well, thus
providing an efficient veto against this kind of events.

\begin{figure}[htb]
 \centering
 \includegraphics[width=.7\textwidth, clip]{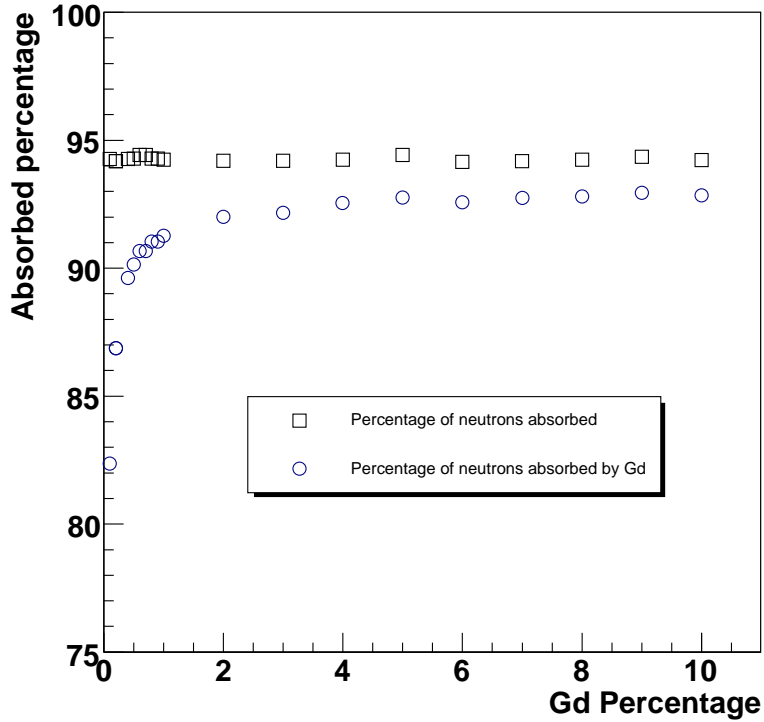}
 \caption{Number of absorbed particles as a function of the Gd
   concentration.}
 \label{fig:gdconc}
\end{figure}

Following the approach discussed in~\cite{gadzooks}, 
we have doped the water-filled parallelepiped 
with highly-soluble gadolinium trichloride (GdCl$_3$). To avoid
the absorption of photons by the cylindrical targets and the 
supporting system associated to them, we
propose a solution similar to the one used in the Pierre Auger
Observatory~\cite{pao}, namely to cover their external walls with Tyvek 
(a material that shows a reflectivity higher 
than 90\% to \v{C}erenkov light~\cite{justus}). A particular source of
concern is the radio-purity of the additive. According to the
estimations given in~\cite{gadzooks} and~\cite{sno}, the potential
background caused by it, especially the alpha particle decays of $^{152}$Gd, is
much smaller than what is expected from the sources considered in
Section~\ref{sec:results}.     
The amount of gadolinium has being chosen in order to minimize
the number of neutrons captured by hydrogen nuclei, since we consider 
the 2.2 MeV gammas coming from this reaction are extremely hard to 
detect with the outer veto. A dedicated GEANT4
simulation has been carried out to study which is the optimal 
Gd concentration. A 1~meter radius
sphere filled with Gd-doped water has been simulated and neutrons
with energies up to 10~MeV have been shot from the
center. Figure~\ref{fig:gdconc} shows the obtained results.

We observe that while the total number of absorbed particles does not change
with Gd concentration, the proportion of Gd-absorbed particles 
does, saturating at a value $\sim2\%$. 
Hence, we will use for our calculations a $2\%$ admixture by mass of 
GdCl$_3$. 

To evaluate the veto efficiency, we follow the gammas produced in the 
8 MeV cascade following neutron capture by Gd. The detectable signal 
corresponds to Compton electrons above Cerenkov threshold. More than 
90$\%$ of these electrons have energies above 3 MeV, with a mean value 
of about 5 MeV~\cite{sno}. Nearly 50$\%$ of the Cerenkov photons 
are detected and only 3$\%$ of them are absorbed by the targets and 
their associated ancillary system. Considering a global detection efficiency 
of 15$\%$ for the simulated PMTs, we obtain a light yield of 6 photo-electrons/MeV. 
Assuming a detection threshold of 3 MeV, our trigger efficiency is $>$95$\%$ 
for this energy and reaches $\sim$ 100$\%$ at 4 MeV~\cite{hosaka}. The overall 
detection efficiency for the 8 MeV gamma cascade is $>$90$\%$ and the energy 
resolution is assumed to be 20$\%$. 

\section{Physics performance}
\label{sec:results}

The estimation of the overall 
background event rate must take into account
both internal and external sources of gamma rays
and neutrons. We conservatively assume that, 
due to instrumental limitations, we cannot detect signals below 
15 (30) keV of true recoil energy in case we use a xenon (argon)
target. On the other hand, we will assume a maximum true recoil energy for
WIMP like events of 50 (100) keV for the xenon (argon) target. We note
that it has been recently suggested that neutrinos can be a source of
background for the next generation of direct-search dark matter
experiments~\cite{Monroe:2007}. However, as shown 
in~\cite{Monroe:2007}, for Ar and Xe targets, this background is 
easily eliminated considering a cut on the true recoil energy 
above 10 keV. Therefore the cuts we impose throughout 
this work on the true recoil energy reduce 
the neutrino contamination to a negligible level. 

\subsection{Contamination from radioactive nuclei, xenon and argon 
isotopes}
\label{sec:isotopes}

For a target made of argon, an important source of background comes 
from the presence of radioactive $^{39}$Ar. This is a beta particle emitter 
with an activity of about 1~Bq per kg of natural argon, which for a
single-volume 1 tonne detector translates into a 1 kHz rate. In our
case, since the target is divided into hundred independent units, the
event rate due to $^{39}$Ar decays does not represent an issue for the
design of the data-acquisition system. In addition, the probability to have a
$^{39}$Ar decay overlapping with a different sort of interaction is
smaller than in the case of a single-volume large-size detector due to the
smaller drift times involved. 

Regarding the possibility of misidentifying $^{39}$Ar signals as WIMPs, 
we should note that $\beta$ particles mainly interact with atomic
electrons, while nuclear recoils deposit their energy through
transfers to screened nuclei~\cite{Lindhard}. This affects the charge
generated by an event (for the same energy is around three times bigger
for electrons), the charge to light ratio, which is bigger for
electrons, and the ratio between the slow and the fast component of
the scintillation light of liquid argon (pulse shape discrimination).

According to our simulations, 
the background due to radioactive nuclei can be reduced 
to a level well below the one expected from neutrons
using the ratio of measured scintillation
light over ionization and pulse shape discrimination~\cite{pulse,warp}.
For a nuclear recoil acceptance of 
50$\%$, the rejection power against backgrounds caused by electromagnetic 
particles is $\sim 5 \cdot 10^{-7}$ for each individual target. This rejection power 
agrees with the results quoted in~\cite{warp} and
\cite{Lippincott:2008}. A further reduction of this kind of background
will come from the use of underground-extracted argon~\cite{Galbiati:2007}. 
Its $^{39}$Ar activity has been recently measured for the first time 
and shown to be $<$5\% of the one present in natural argon. These
reasons lead us to not consider further this sort of background.

In case the target is filled with xenon, $^{136}$Xe is the most important
radioactive isotope. It decays through double beta decay and therefore, given the
small probability of the process, the resulting count rate, in the
energy band of interest, is negligible compared to other 
sources of background, even before any rejection cut is applied.

Krypton and radon are two radioactive nuclides present in
commercially available noble gasses and therefore a potential source
of background as well. The highest contamination 
comes from $^{85}$Kr, which $\beta$--decays with an endpoint energy 
of 678~keV. As has been discussed, impurities of Kr
below 10~ppb can be reached~\cite{xenon10}, making negligible the contamination produced
by those radioactive decays. 


\subsection{Neutrons from target components}

One of the most important sources of background comes from neutrons
produced by radioactive contamination of the materials constituting
the detector itself. To minimize their rate, the use of copper for all the 
vessels is likely to be the best possible choice. The radioactive impurities can be
reduced below 0.02~ppb in some copper samples which would bring the neutron
contamination to below 1 event per year~\cite{ILIAS}. If we
conservatively assume a
0.1~ppb contamination, one obtains a neutron production rate
of $4.54\times10^{-11}$ s$^{-1}$cm$^{-3}$. Being each cylinder 6~mm thick,
its total volume amounts to 2217 cm$^{3}$. This
means a total of one neutron per cylinder per year. 

The contamination induced by PMTs must be carefully evaluated as well. 
Main manufacturers continue to optimize the choice of
materials used in PMT construction to reduce their radioactivity levels.
Typical contamination values for U and Th range from a few tens
to several hundreds parts per billion. 
Among the wide variety of tubes available in the market, it is possible
to find out some models specially designed for low background applications
where the measured uranium and thorium concentrations in quartz and
metal components is of the order of ten or even less
ppbs~\cite{Kudryavtsev}, giving a yearly production of less than one
neutron per PMT. The phototube windows could be coated with
Tetra-Phenyl-Butadiene (TPB) to shift the ultra-violet light to the
maximum of the phototube spectral response without an increase on
contamination. If we assume a rate of 1 neutron emitted per year per PMT
and 8 PMTs per cylinder, we expect a total emission of 
8 neutrons per cylinder. In total, PMTs and the copper vessel
contribute to 9 neutrons emitted per cylinder per year. 

Although they will not be considered in the present work, there are
several possibilities to reduce the rate of neutrons coming from
PMTs. One is to set acrylic light-guides between photomultipliers and
the active volume~\cite{Kudryavtsev} which can reduce by a factor 2
the rate of neutrons. Another possibility is to substitute the
PMTs on the top of the cylinder by charge readout devices, 
which can be constructed from low radioactivity materials, 
having a negligible neutron production rate.

We have studied the background rate due to detector components 
using a simulated sample that amounts to 50 years of data
taking. The results shown in Tables~\ref{tab:detres}
and~\ref{tab:detresXe} are normalized to one year of operation. 
Table~\ref{tab:detres} corresponds to the configuration where LAr is
used. Throughout this work, columns labeled as {\it Total} refer to the
total number of neutrons per year, while columns labeled as {\it Not vetoed}
refer to those neutrons not being absorbed in the Gd-doped water
tank; likewise by $E_{recoil}$ we mean the equivalent recoil energy 
inferred from the energy measured in the active target.

\begin{table}[ht]
  \begin{center}
     \begin{tabular}{|c|c|c|}
       \hline
       Neutrons & Total & Not vetoed\\
       \hline
       Produced in 1 year & 900 &  20 \\
       30 keV $<$ E$_{recoil}<$ 100 keV & 19 & 0.3 \\
       \hline
     \end{tabular}
\end{center}
\caption{LAr target: Neutron background from detector
components normalized to one year of data taking. 
\label{tab:detres}}
\end{table}

After a simple selection cut based on the nuclear recoil energy, 
we find a background of 0.3 neutrons per year for a
LAr detector with a fiducial mass of 0.4 tonnes. Considering as signal only those neutrons
interacting just once inside the active volume, we can get rid 
of some additional background. However given the small dimensions of the targets, we
expect a modest reduction factor from events with multiple interactions. It is important to note that when
the active Gd-doped veto is used, the amount of background is reduced
by roughly a factor fifty. The results using liquid xenon as target (fiducial mass 0.8 tonnes) 
are shown in Table~\ref{tab:detresXe}. The overall background, after the energy
cut, amounts to 1 neutron per year. The reduction given by the
active veto in this case is only a factor ten. The amount of
background for xenon is larger than for argon. The reason comes from the 
fact that
some xenon isotopes like $^{131}$Xe and $^{129}$Xe show a very high
cross section for neutron absorption. For the case of a xenon-filled
detector, the smaller the dimensions of the target cylinder the better to identify neutrons
in the external active veto.

\begin{table}[ht]
  \begin{center}
     \begin{tabular}{|c|c|c|}
       \hline
       Neutrons & Total & Not vetoed\\
       \hline
       Produced in 1 year & 900 &  64 \\
       15 keV $<$ E$_{recoil}<$ 50 keV  & 12 & 1 \\
       \hline
     \end{tabular}
\end{center}
\caption{LXe target: Neutron background from detector components
normalized to one year of data taking. \label{tab:detresXe}}
\end{table}

\subsection{Neutrons and gamma rays from active veto components}
Assuming the same contamination levels we used in the previous 
Section to estimate the neutron flux due to copper walls, PMTs, 
voltage divider bases, etc., we obtain that the active veto 
system contributes with approximately $10^4$ emitted neutrons per year. 
The flux of these neutrons is orders of magnitude smaller 
than the ones that reach the external walls of the 
detector, after being produced in the rock of the cavern 
by natural radioactivity (according to Table~\ref{tab:rockres} it 
amounts to O(10$^7$) rock-emitted neutrons per year). Therefore 
the contribution of the neutron-induced background 
from veto components is added to the contamination 
induced by the walls of the cavern and will be treated 
in the next Section, but it represents a 
small fraction of the total expected background.   
    
Another source of contamination is the gamma ray flux produced 
by the PMTs of the veto system. They mainly come from 
the decay of $^{208}$Tl (thorium chain) 
and $^{214}$Bi (uranium chain). The former produces a 
2.6 MeV gamma and the latter emits 2.2 MeV and 2.4 MeV photons. 
As explained before, we can set a threshold of 3 MeV for the 
veto system without a significant loss of efficiency. 
In these conditions, the majority of those gamma rays 
will fall below threshold. Those reaching the targets 
can be rejected using the criteria discussed in 
Section~\ref{sec:isotopes}, and therefore their contribution 
to the total background will be significantly smaller than 
the one expected from neutrons.

\subsection{ Neutrons from surrounding rock}

 Neutrons coming from the rock have two possible origins:
(1) underground production by cosmic muons (called hereafter 
``muon--induced neutrons'') and (2) neutrons induced by spontaneous 
fission and ($\alpha$, n) reactions due to uranium and thorium present 
in the rock (generically called from now on ``radioactive''). The latter have a
very soft spectrum (typically energies of few MeV). 
The energy spectrum from muon--induced neutrons is harder and 
therefore are more difficult to moderate by the
water shielding. Those neutrons may come from larger distances and produce recoils with 
energies well above the energy threshold set for signal 
events~\cite{Kudryavtsev}. The active external
water veto will efficiently tag crossing muons by \v{C}erenkov light
detection. Neutron signals occurring in the noble liquid target
in coincidence with water PMT signals will be rejected. 
There are $\sim10^{7}$ neutron absorptions per year
in the water volume. If we assume a 100 $\mu$s veto time per interaction,
this will correspond to a total of $\sim20$ min dead time of the
detector per year due to neutron interactions. 

\subsubsection{Neutrons from radioactivity:}

\begin{figure}[htb]
 \centering
 \includegraphics[width=.7\textwidth, clip]{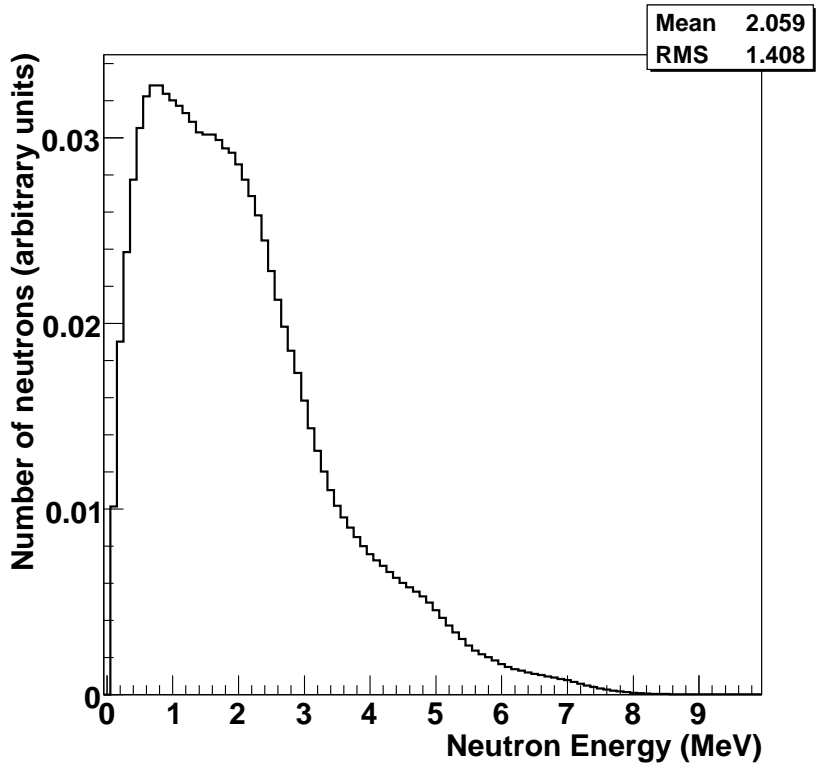}
 \caption{Energy spectrum of neutrons produced in the rock by natural
   radioactivity.}
 \label{fig:enerock}
\end{figure}

Natural radioactivity can produce neutrons either directly from
spontaneous fissions or by means of emitted alpha particles through $(\alpha,
n)$ reactions. To compute the spectrum and the rate of those neutrons, the program
SOURCES-4C~\cite{SOURCES} has been used. Since the original program 
provides $(\alpha,n)$ reactions up to 6.5 MeV $\alpha$ particle energy, 
we include the modifications 
done in~\cite{Kudryavtsev} to obtain a more realistic neutron spectrum. 
The thorium and uranium
contamination has been taken as the average of those given in
reference~\cite{Amare}, namely 18.8 Bq/kg for $^{238}$U and 42 Bq/kg
for $^{232}$Th. Secular equilibrium is considered. 
Accordingly, the computed rates for neutron production
amounts to $8.44\times10^{-8}$ s$^{-1}$cm$^{-3}$ from $(\alpha, n)$ reactions and
to $7.38\times10^{-8}$ s$^{-1}$cm$^{-3}$ from spontaneous
fission. Neutrons have been generated according to the energy
spectrum shown in Figure~\ref{fig:enerock} and propagated through the
rock using the GEANT4 simulation code and the prescriptions given
in~\cite{Kudryavtsev}. As a result we get the neutron 
spectrum in the walls of the laboratory. To get the final number of
neutrons impinging in the detector outer walls and their energy, we
have simulated a cavern of $15\times12\times40$ m$^3$, similar in dimensions to the
experimental main hall at Canfranc underground laboratory~\cite{Canfranc}. 
Table~\ref{tab:rockres} shows the number of neutrons that reach  
the detector and those that produce energy deposits in the
same range as WIMP interactions. Normalized to one year of data
taking, we show the level of expected background for two different
distances between the external vessel wall and the first active
cylinder a neutron will encounter. 
With a 60~cm thick water active veto, the number of interactions
inside the liquid argon volume is well below 1 per year.

\begin{table}
  \begin{center}
     \begin{tabular}{|c|c|c|c|c|}
       \hline
       & \multicolumn{2}{c|} {40~cm of water}& \multicolumn{2}{c|} {60~cm of
       water}\\
       \cline{2-5}
       Neutrons & Total & Not vetoed&Total & Not vetoed\\
       \hline
       Produced in 1 year & $4.2\times10^7$ &  $1.6\times10^7$ &
       $4.2\times10^7$ & $1.6\times10^7$\\
       30~keV$<$$E_{recoil}$$<$100~keV & $175$ & $3$ & $9$ &$<0.1$\\
       \hline
     \end{tabular}
\end{center}
\caption{Neutron background from rock radioactivity. 
We assume a LAr target with a fiducial mass of 0.4 tonnes and one year of data taking. 
Results are shown for two different configurations of 
the active water veto.\label{tab:rockres}} 
\end{table}

The study has been repeated considering a 
liquid xenon target and 60~cm thick water veto. For this configuration, the expected
background from rock radioactivity amounts to nearly one event per year (see
Table~\ref{tab:rockresXe}). In accordance with the results got while studying the contamination due 
to neutrons from detector components, once more the larger cross section for neutron absorption 
is responsible for having a bigger expected background when xenon is considered as detector target.

\begin{table}
  \begin{center}
     \begin{tabular}{|c|c|c|}
       \hline
       Neutrons & Total & Not vetoed\\
       \hline
       Produced in 1 year & $4.2\times10^7$ & $1.6\times10^7$\\
       15~keV$<$$E_{recoil}$$<$50~keV & $5$ & 0.7 \\
       \hline
     \end{tabular}
\end{center}
\caption{Neutron background from rock radioactivity. A LXe target (0.8 tonnes fiducial mass) has
been considered together with an active water veto 60~cm thick. Results are shown for one year of 
data taking\label{tab:rockresXe}} 
\end{table}

\subsubsection{Muon-induced neutrons:}

Fast neutrons from cosmic ray muon interactions 
represent an important background
for dark matter searches. Unlike charged particles, they can not be
tagged by veto systems, and unlike lower energy neutrons from rock
radioactivity, they can not be stopped by a passive shielding. However,
as proposed in~\cite{akerib}, it is possible to place close to the
detector some material in which this fast neutrons produce secondary
low energy neutrons that can be detected by the proposed veto system 
when absorbed by Gd.

The total muon-induced neutron flux $\phi_{n}$ as a function of the depth for a
site with a flat rock overburden can be estimated as~\cite{Mei:2006}:
\begin{center}
\begin{equation}
\phi_{n}=P_0\left(\frac{P_1}{h}\right)e^{-h/P_1}
\end{equation}
\end{center}
where $h$ is the vertical depth in kilometers water equivalent (km.w.e.),
$P_0=4.0\times10^{-7}$ cm$^{-2}$ s$^{-1}$ and $P_1=0.86$ km.w.e.

\begin{figure}[htb]
 \centering
 \includegraphics[width=.7\textwidth, clip]{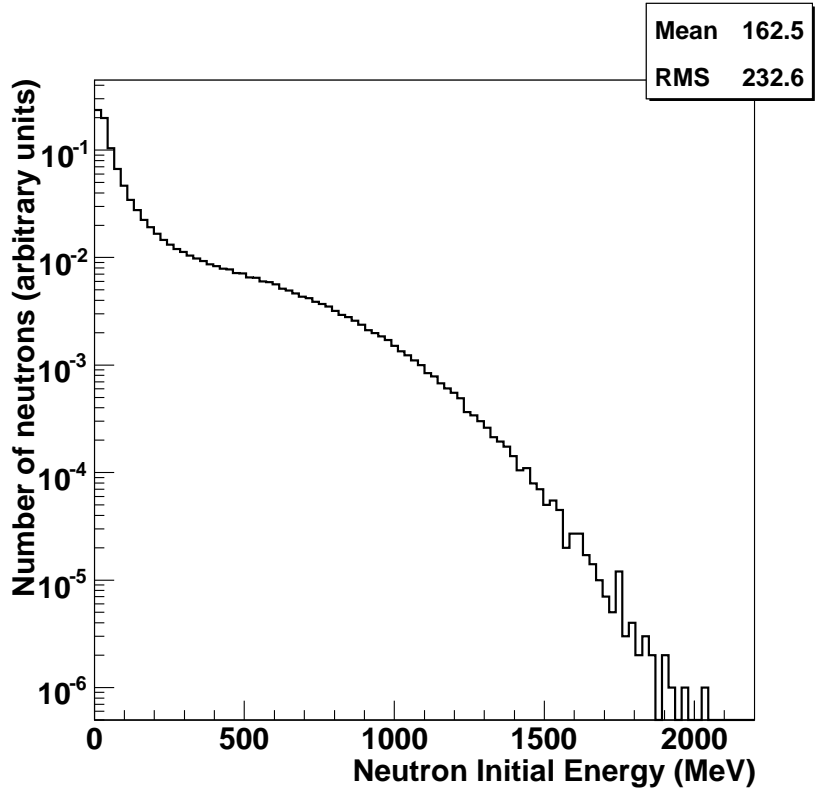}
 \caption{Energy spectrum of neutrons produced by muons interacting in
the surrounding rock as estimated for the Canfranc lab.}
 \label{fig:muonsrock}
\end{figure}

If we consider the Canfranc underground laboratory, with
a depth of $\sim2500$ m.w.e.~\cite{Canfranc}, the total neutron flux is
$7.52\times10^{-9}$ cm$^{-2}$ s$^{-1}$. The neutron energy spectrum is given
by \cite{Wang:2001}:

\begin{center}
\begin{equation}
\frac{dN}{dE_{n}}=A\left(\frac{e^{-7E_{n}}}{E_{n}}+B(E_{\mu})e^{-2E_{n}}\right)
\end{equation}
\end{center}
$A$ is a normalization constant and $B(E_\mu)=0.52-0.58e^{-0.0099E_\mu}$. 
The muon energy spectrum
can be estimated with the following equation~\cite{Mei:2006}:
\begin{center}
\begin{equation}
\frac{dN}{dE_{\mu}}=Ce^{-bh(\gamma_\mu-1)} \cdot \left(E_\mu+
\epsilon_\mu(1-e^{-bh}) \right)^{-\gamma_\mu}
\end{equation}
\end{center}
where $C$ is a normalization constant, $E_\mu$ is the muon energy in GeV,
$b=0.4$/km.w.e, $\gamma_\mu=3.77$ and
$\epsilon_\mu=693$ GeV. The previous equations give rise to the energy
spectrum displayed in Figure~\ref{fig:muonsrock}.

The angular neutron distribution can be expressed as~\cite{Wang:2001}:
\begin{center}
\begin{equation}
\frac{dN}{dcos\theta}=\frac{A}{(1-cos\theta)^{0.6}+B(E_\mu)}
\end{equation}
\end{center}
with $B(E_\mu)=0.699E_\mu^{-0.136}$.

In order to simulate the fast neutron background, we consider a
$10\times10$ m$^2$ surface on the detector from which we simulate neutrons
with the specified angular and energy distributions. Together with
the detector itself, we simulate a lead block in which neutrons will
create secondary particles.  We have considered two different 
configurations (lead block on the top or at the
bottom of the detector) and two different thicknesses 
for the passive lead veto. 

\begin{table} 
 \begin{center}
\begin{tabular}{|c|cc|cc|}
       \hline
 Neutrons & \multicolumn{2}{c}{Total} & \multicolumn{2}{|c|}{Not vetoed} \\\hline
Produced in 1 year & \multicolumn{2}{|c}{$2.4\cdot10^5$} & \multicolumn{2}{|c|}{$1.9\cdot10^5$} \\
\hline 
\multicolumn{5}{|c|}{{\bf No lead}}\\\hline
30~keV$<$$E_{recoil}$$<$100~keV & \multicolumn{2}{|c}{145} & \multicolumn{2}{|c|}{22}\\
Muon veto & \multicolumn{2}{|c}{15} & \multicolumn{2}{|c|}{2}\\
       \hline
\multicolumn{5}{|c|}{{\bf Lead block (bottom)}}\\\hline
& \underline{30 cm} & \underline{60 cm}& \underline{30 cm}& \underline{60 cm} \\
30~keV$<$$E_{recoil}$$<$100~keV & 149 & 149 & 8 & 11\\
Muon veto & 15 & 15 & 0.7 & 1 \\
       \hline 
\multicolumn{5}{|c|}{{\bf Lead block (top)}}\\\hline
& \underline{30 cm} & \underline{60 cm}& \underline{30 cm}& \underline{60 cm} \\
30~keV$<$$E_{recoil}$$<$100~keV & 34 & 9 & 4 & 0.3\\
Muon veto & 3 & 1 & 0.4 & $<0.1$ \\
       \hline
     \end{tabular}
\end{center}
\caption{Background events coming from cosmic muon-induced neutrons
  for different LAr detector configuration: a) no additional passive veto; 
b) an additional passive veto, located either on top or at the bottom of the detector, 
made of a 30 (60) cm thick lead block.We assume one year of data taking.\label{tab:muonres}}
\end{table}

According to Table~\ref{tab:muonres}, out of the four configurations studied, 
the best one corresponds to the 
case where a 60~cm thick lead block is placed on top of
the detector. With this passive veto alone, the background is about ten events 
per year, provided the lead is 60 cm thick. When combined with the Gd-doped water 
veto, the background drops well below one event per year. The simulations have been repeated
considering liquid xenon as the target material. Results are shown in
Table~\ref{tab:muonresXe}. Again, the expected background coming from 
muon-induced neutrons is
$\sim$1 event per year for the whole detector. However, further reduction can be 
achieved when water itself is considered as an active veto 
(as shown in Tables~\ref{tab:muonres} and~\ref{tab:muonresXe}, 
where we refer to it as Muon veto). 
It has been demonstrated that by rejecting events in
coincidence with a muon, the contamination level decreases by a
factor 10~\cite{Kudryavtsev}. In our case, this means 
the overall background would be well below 1 event per year per tonne 
of target material.

\begin{table} 
 \begin{center}
     \begin{tabular}{|c|c|c|}
       \hline
       Neutrons & Total & Not vetoed\\
       \hline
       Produced in 1 year & $2.4\cdot10^5$ &  $1.9\cdot10^5$\\
       15~keV$<$$E_{recoil}$$<$50~keV& 8 & 0.7 \\
       Muon veto & 0.8 & $<0.1$\\
       \hline
     \end{tabular}
\end{center}
\caption{Background events coming from cosmic muon-induced neutrons
  using LXe as target material for a data taking period of one year. 
We assume that a 60 cm thick 
lead block is installed on top of the detector.\label{tab:muonresXe}}
\end{table}

\section{Discussion} 

\begin{figure}[htb]
 \centering
 \includegraphics[width=.7\textwidth, clip]{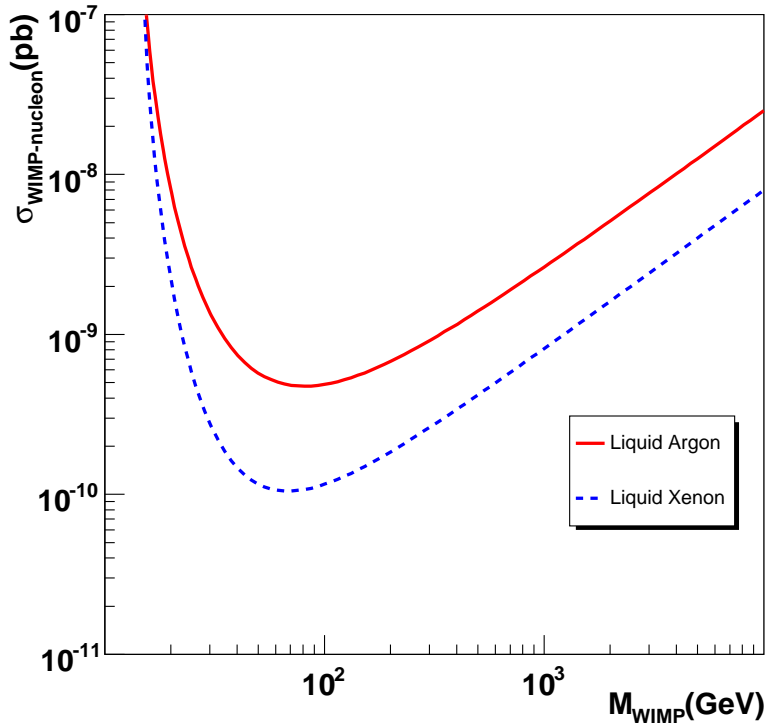}
 \caption{Achievable sensitivities for the case sensitive targets are 
filled either with LAr or LXe. The curves have been computed assuming 
an exposure of one tonne $\times$ year, 50$\%$ nuclear recoil acceptance. 
The tool from reference has 
been used~\cite{DAMNED}.}
 \label{fig:limit}
\end{figure}

As a result of our study, we have seen that 
the combination of a noble liquid (used as sensitive target) 
and a Gd-doped active water veto efficiently reduces neutron background.
For idealized data taking conditions, if we take as reference value an exposure of one 
tonne $\times$ year, the total neutron-induced contamination 
for the case of an argon-filled detector is one
event, while two events are expected for the case of xenon (see Table~\ref{tab:summary}).
As shown in Figure~\ref{fig:limit}, in case no statistically significant signal is 
observed, we can reach 
sensitivities~\cite{Feldman} for the WIMP-nucleon spin-independent
cross section close to 10$^{-10}$ pb. To compute these limits we have assumed 
a standard dark matter galatic halo~\cite{lewin}, 
an energy resolution that amounts to 25$\%$ for the energy range of 
interest and 50$\%$ nuclear recoil acceptance. 
For completeness, we note that 
the best upper limit to date excludes (at 90$\%$ C.L.) cross sections 
above $4.5 (4.6)\times 10^{-8}$ pb for a WIMP mass of 30(60) GeV/c$^2$~\cite{xenon10,cdmslimit}.   

\begin{table} 
 \begin{center}
     \begin{tabular}{|c|c|c|}
       \hline
       Background & LAr target & LXe target\\
        & (exposure: 1 tonne $\times$ year) & (exposure: 1 tonne $\times$ year)\\
       \hline
       Detector components & 0.9 &  1.2 \\
       Rock radioactivity & $\!\!\!\!<$0.1 & 0.9 \\
       Muon-induced &  0.1 &  0.1 \\
       {\bf Total}& {\bf 1.0} & {\bf 2.2} \\
       \hline
     \end{tabular}
\end{center}
\caption{Total expected backgrounds for two different target configurations. 
Figures have been normalized to 
an exposure of one tonne per year.\label{tab:summary}}
\end{table}

\section{Conclusions} 

We have evaluated the performance of a detector devised
to carry out a direct search for WIMPs. It is made of 
two sub-detectors: the active target consists 
of small cylinders filled with a liquefied
noble gas. They are immersed inside an ultra-pure water tank doped with 
gadolinium, that acts as a veto system against neutrons and cosmic muons. 
This configuration enhances the probability for neutron capture in water
and its identification, thus providing a much improved rejection tool 
against this kind of background. This technique is scalable and allows
the construction of large detectors with masses in the tonne range.

We have observed that the use of a Gd-doped veto reduces by 
about a factor fifty the neutron contamination in case the target is filled
with argon and up to a factor ten in case xenon is used. 

In case a positive WIMP signal is observed with sufficient
statistical power, we can confirm, with a single experiment, that 
the event rate and the recoil spectral shape follow 
the expected dependence on A$^2$, since the independent target units
can be filled with different noble liquids. 

A simulation of
the potential background sources has shown that  for 
an exposure of one tonne $\times$ year, we expect 
a contamination of about one event. If no WIMP signal is observed, 
our calculation shows that, for idealized data taking conditions, this exposure 
will suffice to exclude spin-independent WIMP-nucleon cross sections in the 
range $10^{-9} - 10^{-10}$ pb. 

\section*{Acknowledgments}

We thank S. Navas for useful comments and ideas while reading
this manuscript. This work has been done under the auspices of 
M.E.C. (Grant FPA2006-00684).




\section*{References}
\begin{thebibliography}{99}
\bibitem{Zwicky} Zwicky F, 1933 {\it Hel. Phys. Acta} {\bf 6} 110.
\bibitem{Trimble} Trimble V, 1987 {\it Ann. Rev. Astron. Astrophys.} {\bf 25} 425.
\bibitem{bullet} Clowe, D. {\it et al.}, 2006 {\it Astrophys. J.} {\bf 648} L109-L113.
\bibitem{bullet2} Bradac, M. {\it et al.}, 2006 {\it Astrophys. J.}
{\bf 652} 937.
\bibitem{Sanders} Sanders R H and Mc Gaugh S S, 2002 {\it 
Ann. Rev. Astron. Astrophys.} {\bf 40} 263.
\bibitem{wmap} Tegmark M {\it et al.}, 2006 {\it Phys. Rev. D} {\bf 74} 123507.
\bibitem{atm} Fukuda Y {\it et al.}, 1998 {\it Phys. Rev. Lett.} {\bf 81} 1562.
\bibitem{solar} Ahmad Q R {\it et al.}, 2002 {\it Phys. Rev. Lett.} {\bf 89} 011301.
\bibitem{Bertone} Bertone G, Hooper D and Silk J, 2005 {\it Phys. Rept.} {\bf 405} 279.
\bibitem{Gaitskell} Gaitskell R J, 2004 {\it
Ann. Rev. Nucl. Part. Sci.} {\bf 54} 315.
\bibitem{Bernabei:2000qi}
Bernabei R {\it et al.},
2000 {\it Phys. Lett. B} {\bf 480} 23; 2003 {\it Riv. N. Cim.} {\bf 26} 1.
\bibitem{Bernabei:2008yi}
  Bernabei R {\it et al.},
  2008 arXiv:0804.2741.
\bibitem{Akerib:2004fq}
Akerib D S {\it et al.},
2004 {\it Phys. Rev. Lett.} {\bf 93} 211301.
\bibitem{Chardin:2003vn}
Chardin G {\it et al.},
2004 {\it Nucl. Instrum. Meth. A} {\bf 520} 101.
\bibitem{Ahmed:2003su}
Ahmed B {\it et al.},
2003 {\it Astropart. Phys.}  {\bf 19} 691.
\bibitem{warp} Benetti P {\it et al.}, 2008 {\it Astropart. Phys.} {\bf 28} 495.
\bibitem{xenon10} Angle J {\it et al.}, 2008 {\it Phys. Rev. Lett.}
{\bf 100} 021303.
\bibitem{Ambrosio:1998qj}
Ambrosio M {\it et al.},
1999 {\it Phys. Rev. D} {\bf 60} 082002.
\bibitem{Desai:2004pq}
Desai S {\it et al.},
2004 {\it Phys. Rev. D} {\bf 70} 083523.
\bibitem{nobleliquids1}
Belli, P {\it et al.}, 1990 {\it N. Cim. A} {\bf 103} 767.
\bibitem{nobleliquids2}
Benetti, P {\it et al.}, 1993 {\it Nucl. Instrum. Meth. A} {\bf 327} 203.
\bibitem{nobleliquids3}
Davies, G J {\it et al.}, 1994 {\it Phys. Lett. B} {\bf 320} 395.
\bibitem{andre} Laffranchi M and Rubbia A, arXiv:hep-ph/0702080.
\bibitem{pfsmith} Smith P F, 2005 {\it New Astronomy Reviews} {\bf 49} 303.
\bibitem{rusos} Dolgoshein B A, Lebedenko V N and Rodionov B U, 1970 {\it JETP Lett.} {\bf 11} 513.
\bibitem{Hitachi} Hitachi A. {\it et al.}, 1983 {\it Phys. Rev. B}
{\bf 27} 5279.
\bibitem{zeplin1}
Alner, G J {\it et al.}, 2007 {\it Astrop. Phys.} {\bf 28} 287. 
\bibitem{zeplin2}
Alner, G J {\it et al.}, 2007 {\it Phys.Lett. B} {\bf 653} 161.
\bibitem{zeplin3}
Araujo, H M {\it et al.}, 2006 {\it Astrop. Phys.} {\bf 26} 140. 
\bibitem{zeplin4}
Akimov, D Y {\it et al.}, 2007 {\it Astrop. Phys.} {\bf 27} 46.
\bibitem{xmass}
Kim, Y D {\it et al.}, 2006 {\it Phys.Atom.Nucl.} {\bf 69} 1970.
\bibitem{betapaper} 
Bueno A, Carmona M C, Lozano J and Navas S, 2006 {\it Phys. Rev. D}
{\bf 74} 033010.
\bibitem{Ar39radio}
Benetti P {\it et al.}, 2007 {\it Nucl. Instrum. Meth. A} {\bf 574} 83.
\bibitem{cline}
Bungau C {\it et al.}, 2005 {\it Astropart. Phys.} {\bf 23} 97.
\bibitem{akerib} 
Hennings-Yeomans R and Akerib D S, 2007 {\it Nucl. Instrum. Meth. A} {\bf 574} 89.
\bibitem{gadzooks} 
Beacom J F and Vagins M R, 2004 {\it Phys. Rev. Lett.} {\bf 93} 171101.
\bibitem{Allison}
Allison J {\it et al.}, 2006 {\it IEEE Trans. Nucl. Sci.} {\bf 53} 270.
\bibitem{Sauli}
Sauli F, 1997 {\it Nucl. Instrum And Meth. A} {\bf 386} 531.
\bibitem{Jeanneret}
Jeanneret P {\it et al.}, 2003 {\it Nucl. Instrum And Meth. A} {\bf 500} 133.
\bibitem{Giomataris2}
Giomataris Y {\it et al.},
1996 {\it Nucl. Instrum And Meth. A} {\bf 376} 29.
\bibitem{pao}
Abraham J {\it et al.}, 2004 {\it Nucl. Instrum And Meth. A} {\bf 523} 50.
\bibitem{justus} 
Gichaba J O, Master's Thesis, University of Mississippi, 1998.
\bibitem{sno} 
Hargrove C K {\it et al.}, 1995 {\it Nucl. Instrum And Meth. A} {\bf 357} 157.
\bibitem{hosaka}
Hosaka J {\it et al.}, 2006 {\it Phys. Rev. D} {\bf 73} 112001.
\bibitem{Monroe:2007}
  Monroe J and Fisher P,
  2007 {\it Phys. Rev. D} {\bf 76} 033007.
\bibitem{Lindhard} Lindhard J {\it et al.},  1963 {\it
  Mat. Fys. Medd. Dan. Vid. Selsk.} {\bf 33} 10. 
\bibitem{pulse} Alner, G J {\it et al.}, 2005 {\it Astrop. Phys.} {\bf 23} 444.
\bibitem{Lippincott:2008} Lippincott W H {\it et al.}, arXiv:0801.1531.
\bibitem{Galbiati:2007} Galbiati C {\it et al.}, arXiv:0712.0381.
\bibitem{ILIAS} ILIAS database on radiopurity materials
  http://radiopurity.in2p3.fr/
\bibitem{Kudryavtsev}
Carson M J {\it et al.}, 2004 {\it Astropart. Phys.} {\bf 21} 667.
\bibitem{SOURCES} Wilson W B {\it et al.} SOURCES-4A, Technical Report
  LA-13639-MS, Los Alamos (1999).
\bibitem{Amare} Amare J {\it et al.},2006 {\it J. Phys.: Conf. Ser.}
{\bf 39} 151. 
\bibitem{Canfranc} Canfranc Underground Laboratory site,
  http://ezpc00.unizar.es/lsc/index2.html
\bibitem{Mei:2006} Mei D M and Hime A, 2006 {\it Phys. Rev. D} {\bf
  73} 053004.
\bibitem{Wang:2001} Wang Y F {\it et al.}, 2001 {\it Phys. Rev. D}
  {\bf 64} 013012.
\bibitem{Feldman}
  Feldman G J and Cousins R D,
  1998 {\it Phys. Rev. D} {\bf 57} 3873.
\bibitem{lewin} Lewin J D and Smith P F, 1996 {\it Astropart. Phys.} {\bf 6} 87.
\bibitem{cdmslimit} Ahmed Z {\it et al.}, arXiv:0802.3530.
\bibitem{DAMNED} http://pisrv0.pit.physik.uni-tuebingen.de/darkmatter/limits/index.php
\end{thebibliography}
\end{document}